# Restarting Automata with Auxiliary Symbols and Small Lookahead[*]

Natalie Schluter

IT University of Copenhagen
Rued Langgaards Vej 7, 2300 Copenhagen S., Denmark
`nael@itu.dk`

**Abstract**

We present a study on lookahead hierarchies for restarting automata with auxiliary symbols and small lookahead. In particular, we show that there are just two different classes of languages recognised by RRWW automata, through the restriction of lookahead size. We also show that the respective (left-) monotone restarting automaton models characterise the context-free languages and that the respective right-left-monotone restarting automata characterise the linear languages both with just lookahead length 2.

## 1 Introduction

Restarting automata work in phases of scanning their input from the left end marker towards the right end marker, rewriting the lookahead contents with a shorter substring once per phase, and then restarting at some point before or at the right end marker. They were introduced to model the analysis by reduction grammar verification technique in the analysis of sentences in free-word order natural language. It has been shown that through various restrictions on the model, an important number of traditional and new formal language classes may be defined. Study of restarting automata has therefore also become important for both its original intent of computational linguistic application development, as well as for being an alternative machine model for investigating properties of traditional and newly distinguished formal language classes.

In his study of lookahead hierarchies, Mraz [3] showed that the expressive power of restarting automata without auxiliary symbols increases with the size of the lookahead. Schluter [6] later showed that for deterministic monotone and monotone restarting automata with auxiliary symbols, separation of rewrite and restart step is not a significant restriction on expressive power for any fixed lookahead size $k \geq 3$, and that for the deterministic model, the difference in power of the models can be overcome by approximately doubling the lookahead size, when $k \geq 3$. In both studies, it was remarked that lookahead hierarchies collapse for (left-)mon-RWW and (left-)mon-RRWW automata to $k = 3$. This paper presents a study on lookahead hierarchies for $k < 3$ of restarting automata with auxiliary symbols. In doing so, we also establish lookahead hierarchies for the most general model of restarting automata, for any $k$. In particular, we show that there are only two different classes of languages recognised by RRWW automata, through restrictions on lookahead size.

We also partially improve a result from [6] and [3], by showing that the respective monotone and left-monotone restarting automaton models characterise the context-free languages with only lookahead size 2. And, we establish a corresponding result for the characterisation of the linear languages by the respective right-left-monotone restarting automata with lookahead size 2.

Following the definition of restarting automata and presentation of some useful properties in Section 2, we present our main results in Section 3.

---

[*]This is the full version of the paper accepted at LATA 2011.

**Some notation.** We refer to the $i$th symbol of a string $x$ as $x[i]$, and its substring from the $i$th to $j$th symbols as $x[i,j]$. When we want to make the length of a string $v$ such that $|v| = k$ explicit, we may refer to $v$ as $v[1,k]$.

For $i, j \in \mathbb{N}$, with $i < j$, $[i,j]$ alone denotes the set $\{i, \ldots, j\}$. If $i = 1$, we say $[j] := [1, j]$.

If $S$ is a set of symbols, then by $S^i$ we denote the set of strings of length $i \in \mathbb{N}$ with symbols from $S$. Also $\lambda := S^0$ is the empty string.

Finally, REG, LIN and CFL denote the classes of regular, linear, and context-free languages respectively.

## 2 Preliminaries

A restarting automaton or RRWW-automaton, $M = (Q, \Sigma, \Gamma, ¢, \$, q_0, k, \delta)$, is a nondeterministic machine model with a finite control unit and a lookahead (or read/write) window of size $k$ (including the symbol under its scanning head, which is the first symbol of the lookahead contents) that works on a list of symbols delimited by end markers (or sentinels) ($\{¢, \$\}$), where $¢$ is the left sentinel and $\$$ is the right sentinel. $\Sigma$ is the input alphabet and $\Gamma \supseteq \Sigma$ the work tape alphabet. The symbols $\Gamma - \Sigma$ are called *auxiliary symbols*. $Q$ is the finite set of states and $q_0 \in Q$ is the initial state.

$M$'s transition relation, $\delta$, describes four types of transition steps (or instructions), where $u$ is the contents of the lookahead.

(1) A *move-right step* is of the form $q' \in \delta(q, u)$, where $q, q' \in Q$. This means that $M$ advances one tape square to the right and enters state $q'$ upon reading $u$.

(2) A *rewrite step* is of the form $(q', \texttt{REWRITE}(v)) \in \delta(q, u)$, where $q, q' \in Q$, and $v$ is such that $|v| < |u|$ ($u, v \in \Gamma^*$). This means that $M$ replaces its window contents $u$ with $v$, advances to the tape square directly to the right of $v$, and enters state $q'$. In this rewrite instruction, we will refer to $u$ as the *redex* and $v$ as the *reduct*.

(3) A *restart step* is of the form $\texttt{RESTART} \in \delta(q, u)$, where $q \in Q$, in which $M$ moves its read/write window to the beginning of the input and enters the initial state.

(4) An *accept step* is of the form $\texttt{ACCEPT} \in \delta(q, u)$, in which $M$ halts and accepts. (This may also be viewed as the accept state.)

If $\delta(q, u) = \emptyset$, in which case we say that $\delta$ is *undefined*, $M$ halts and rejects; we could exclude this possibility through the use of a model with both accept and reject states, in which case all possibilities for $\delta$ are defined. If $|\delta(q, u)| \leq 1$ for all $q, u$, then the restarting automaton is *deterministic*.

A *configuration* of $M$ is $uqv$, where $u \in \{\lambda\} \cup \{¢\} \cdot \Gamma^*$ is the contents of the worktape from the left sentinel to the position of the head, $q \in Q$ is the current state and $v \in \{¢, \lambda\} \cdot \Gamma^* \cdot \{\$, \lambda\}$ is the contents of the worktape from the current first symbol under the scanning head to the right sentinel, and $uv$ is the current contents of the worktape. The head scans the first $k$ symbols of $v$ (or all of $v$ when $|v| \leq k$). A *restarting configuration*, for a word $w \in \Gamma^*$, is of the form $q_0 ¢ w \$$. If $w \in \Sigma^*$, $q_0 ¢ w \$$ is an *initial configuration*. An *accepting configuration* is a configuration with an accepting state.

A *computation* of $M$ for an input word $w \in \Sigma^*$ is a sequence of configurations starting with an initial configuration, where two consecutive configurations are in the relation $\vdash_M$ induced by a finite set of instructions of one of the above mentioned types. The transitive closure of $\vdash_M$ is denoted $\vdash_M^*$. A *phase* of a computation begins with a restarting configuration and (exclusively) either (1) ends with the next encountered restarting configuration, in which case it includes exactly one rewrite step and is called a *cycle*, or (2) halts, in which case it includes at most one rewrite step and is called a *tail phase*. We refer to segments of a computation within a single phase before (resp. after) a rewrite as *left* (resp. *right*) *computation*.



An input word $w$ is *accepted* or *recognised by* $M$ if there is a computation which starts on the initial configuration and finishes in an accepting configuration. Also, we define $\mathcal{L}(M)$ as the language recognised by $M$.

Consider a cycle $C$ and say the configuration from which $M$ carries out a rewrite step is $uqv$ in $C$; we define to the *right distance* of $C$ as $D_r(C) := |v|$ and the left distance as $D_l(C) := |u|$. Let $\mathcal{C} = C_1, C_2, \ldots, C_n$ be a sequence of cycles of a restarting automaton $M$ that, together with possibly a (final) tail phase, are $M$'s computation on some input. If $D_r(C_i) \geq D_r(C_{i+1})$ for all $i \in [n-1]$, we say that $\mathcal{C}$ is *right-monotone* or simply *monotone*. Similarly, if $D_l(C_i) \geq D_l(C_{i+1})$ for all $i \in [n-1]$, we say that $\mathcal{C}$ is *left-monotone*. If $\mathcal{C}$ is both right- and left-monotone, then we say that $\mathcal{C}$ is *right-left-monotone*. If all the sequences of cycles corresponding to computations of a restarting automaton $M$ are monotone (respectively left-monotone, right-left-monotone) then we say that $M$ is *monotone* (respectively *left-monotone*, *right-left-monotone*). We denote the class of monotone RRWW-automata (respectively *left-* or *right-left*-RRWW automata), *mon*-RRWW (*left-mon*-RRWW or *right-left-mon*-RRWW).

Through restrictions on the restarting automaton model, we obtain many types of restarting automata. For instance, RRW-automata are RRWW-automata with no auxiliary symbols ($\Gamma = \Sigma$). An RR-automaton is an RRW-automaton with rewrite instructions that can only delete symbols. An RWW-automaton is an RRWW-automaton, which restarts immediately after any rewrite instruction, and an RW-automaton is an RRW-automaton that restarts immediately after any rewrite instruction. Finally, an R-automaton is an RR-automaton that restarts after any rewrite instruction.

When the rewrite and restart steps are not separated, instead of items (2) and (3) in the description of $\delta$ above, we have simply the following type of instruction.

(2/3) A *rewrite step* (which is combined with restarting) is of the form $\texttt{REWRITE}(v) \in \delta(q, u)$, where $q, q' \in Q$, and $v$ is such that $|v| < |u|$ ($u, v \in \Gamma$). This means that $M$ replaces its window contents $u$ with $v$ and then moves its read/write window to the beginning of the input and enters the initial state.

All notions of monotonicity and determinism and corresponding notation extend to these more restrictive versions in the obvious way.

An $X$ automaton, $X \in \{R, RR, RW, RWW, RRW, RRWW\}$, with lookahead size $k$, will be denoted by $X(k)$. For example, an RRWW(k) automaton is an RRWW automaton with lookahead size $k$.

### 2.1 Restarting Automaton Specification by Regular Constraints

Niemann and Otto [4] describe the behaviour of a non-deterministic restarting automaton $M$ by means of a finite set of *meta-instructions* of the form $(E_1, u \to v, E_2)$ (called *cycle meta-instructions*) and $(E, \texttt{ACCEPT})$ (called *tail meta-instructions*). In these meta-instructions, $E_1, E_2,$ and $E$ are regular languages, which are called the *regular constraints* of the meta-instruction, and $u$ and $v$ are strings such that $u \to v$ stands for a rewrite step of $M$, where $u$ is the redex and $v$ is the reduct. These meta-instructions are applied as follows. In a restarting configuration $q_0 \text{¢} w \$$, $M$ nondeterministically chooses a meta-instruction, say $(E_1, u \to v, E_2)$. Now, if $w$ does not admit a factorisation of the form $w = w_1 u w_2$ such that $\text{¢} w_1 \in E_1$ and $w_2 \$ \in E_2$, then $M$ halts and rejects. Otherwise, one such factorisation is chosen nondeterministically, and $q_0 \text{¢} w \$$ is transformed into the restarting configuration $q_0 \text{¢} w_1 v w_2 \$$. If $(E, \texttt{ACCEPT})$ is chosen, then $M$ halts and accepts, if $\text{¢} w \$ \in E$, otherwise, $M$ halts and rejects. Similarly, the behaviour of an RWW-automaton $M$ can be described through a finite sequence of meta-instructions of the form $(E, u \to v)$ and $(E, \texttt{ACCEPT})$.

### 2.2 Four Useful Properties

This section presents four basic lemmata used in the proofs of the main results in Section 3.

The correctness preserving property is a fundamental property of restarting automata.

**Proposition 1** (Correctness Preserving Property [5]). *Let $M$ be a restarting automaton, and $u, v$ be arbitrary input words from $\Sigma^*$. If $u \in \mathcal{L}(M)$ and $u \vdash_M^* v$ is an initial segment of an accepting computation of $M$, then $v \in \mathcal{L}(M)$.*

It will be useful to simplify the computations of the restarting automata that we discuss (without reducing their power). The next three lemmata serve this purpose.

A nondeterministic restarting automaton $M = (Q, \Sigma, \Gamma, \text{\textcent}, \$, q_0, k, \delta)$ is in *RR-semidet-form* if (1) halting (and restarting for automata with separate rewrite and restart steps) occurs only when the right sentinel is under the lookahead, and (2) move-right steps are deterministic. The following lemma shows that non-deterministic restarting automata with lookahead length $k$ can be assumed w.l.o.g. to be (1) in RR-semidet-form and (2) making move-right steps based only on the first symbol under the lookahead.[1]

**Lemma 2.** *For any $X$-$Y$ automaton, $M_1 = (Q, \Sigma, \Gamma, \text{\textcent}, \$, q_0, k, \delta)$, where $X \in \{$(right-left-, left-)mon-$, \lambda\}$ and $Y \in \{R, RR, RW, RRW, RWW, RRWW\}$, there is $X$-$Y$ automaton, $M_2 = (Q', \Sigma, \Gamma, \text{\textcent}, \$, q_0', k, \delta')$, such that*

1. *$M_2$ is in RR-semidet form,*

2. *$M_2$ makes move-right steps based on the couple $(u[1], q)$, where $u[1]$ is the first symbol under the lookahead and $q$ is $M_2$'s current state,*

*and $\mathcal{L}(M_1) = \mathcal{L}(M_2)$.*

*Proof.* Jančar [1] showed (1). (2) is easily seen by the specification of non-deterministic restarting automata by means of regular constraints. A restarting automaton specified by regular constraints can easily be assured to be in RR-semidet-form. Halting (and restarting for automata with separate rewrite and restart steps) can be made to occur after verification that the tape contents can be factorised according to the selected meta-instruction and once the automaton reaches the right sentinel. Moreover, move-right steps verify membership in a regular language, so not only can these move-right steps be determinised, but they can be determinised based on just the first symbol under the lookahead. Any monotonicity is preserved. □

If a restarting automaton $M$ only rewrites when the contents of its lookahead is full, we say that $M$ has *fixed rewrite size*.

**Lemma 3.** *For any $X$-$Y$ automaton, $M_1$, where $X \in \{$(right-left-, left-)mon-$, \lambda\}$ and $Y \in \{R, RR, RW, RRW, RWW, RRWW\}$, there exists an $X$-$Y$ automaton, $M_2$, that has fixed rewrite size, such that $\mathcal{L}(M_1) = \mathcal{L}(M_2)$.*

*Proof.* For the proof, we construct a restarting automaton $M_2$ from $M_1$ that never rewrites when its lookahead contains less than $k$ symbols (where $k$ is the length of the lookahead), supposing without loss of generality that $M_1$ is in RR-semidet form. We describe the case where restart and rewrite steps are separated, the other case being easily understood from this.

$M_1$'s lookahead can only contain less than $k$ symbols if it also contains the right sentinel. We rely on a simple speed-up of $M_1$'s steps for the cases (1) where the left sentinel is also contained in the lookahead, (2) of a right computation, or (3) of a tail phase.

Otherwise, $M_1$ (with transition relation $\delta_1$) has a rewrite of the form $(p, \texttt{REWRITE}(v\$)) \in \delta_1(q, u\$)$ where $|u\$| < k$. In this case, we "plug up" the rewrite from the left with all strings $\alpha \in \Gamma^{k-|u\$|}$,

---
[1] Here, the decision whether or not to move-right remains non-deterministic; however, the decision of which move-right step to carry out becomes deterministic.

such that $M_1$ from state $q'$ reads $\alpha u\$$ and enters state $q$ with $u\$$ the prefix of its lookahead, giving $(p, \texttt{REWRITE}(\alpha v\$)) \in \delta_2(q', \alpha u\$)$, where $\delta_2$ is $M_2$'s transition relation.

Clearly $\mathcal{L}(M_1) = \mathcal{L}(M_2)$. Also, monotonicity is clearly preserved. □

**Lemma 4.** *For any X-Y automaton, $M_1$, where $X \in \{$(right-left-, left-)mon-,$\lambda\}$ and $Y \in \{RWW, RRWW\}$, with lookahead size $k$, there exists an X-Y automaton, $M_2$, with lookahead size $k$, that reduces its input by only one symbol per cycle, and is such that $\mathcal{L}(M_1) = \mathcal{L}(M_2)$.*

*Proof.* Let $M_1 = (Q, \Sigma, \Gamma, \text{\textcent}, \$, q_0, k, \delta_1)$ be an X-RRWW automaton where $X \in \{$(right-left-, left-)mon-,$\lambda\}$, with fixed rewrite size, in the RR-semi-det form, and that carries out move-right steps based on only the first symbol under the lookahead. Let B be a symbol not in $\Gamma$, which we call the *blank symbol*. We construct $M_2 = (Q \cup \overline{Q} \cup \hat{Q}, \Sigma, \Gamma \cup \{\text{B}\}, \text{\textcent}, \$, q_0, k, \delta_2)$, such that $\mathcal{L}(M_1) = \mathcal{L}(M_2)$, from $M_1$.

In what follows,
$$q, q', p, p' \in Q, \quad u \in (\Gamma \cup \{\text{\textcent}\}) \cdot \Gamma^{k-2} \cdot (\Gamma \cup \{\$\}),$$
$$x \in (\Gamma \cup \{B\})^{k-2} \cdot (\Gamma \cup \{B, \$\}), \quad \text{and} \quad x_1 x_2 \in (\Gamma \cup \{B\})^{k-2}.$$

$M_2$'s state set includes $M_1$'s state set ($Q$), *marked states* for indicating a guess that there are blank symbols on the tape (in left computations) $\overline{Q} := \{\bar{q} \mid q \in Q\}$, and *hat states* for indicating that $M_2$ is working in a right computation, $\hat{Q} := \{\hat{q} \mid q \in Q\}$.

In a restarting configuration, $M_2$ can either rewrite or move-right. Say $M_2$ wants to simulate a move-right step of $M_1$. $M_2$ first guesses whether there are any blank symbols currently on its tape. If $M_2$ guesses that there are blank symbols on it's tape, then it will move into a marked state. Otherwise it will remain in a state from $Q$. So, if $q' \in \delta_1(q_0, u)$, then $M_2$ has both of the following move-right instructions

$$\overline{q'} \in \delta_2(q_0, u) \quad \text{for guesses that there are blank symbols on the tape, and} \tag{1}$$

$$q' \in \delta_2(q_0, u) \quad \text{for guesses that there are no blank symbols on the tape.} \tag{2}$$

For rewrites, if $(p, \texttt{REWRITE}(v)) \in \delta_1(q, u)$, then

$$(\hat{p}, \texttt{REWRITE}(\text{B}^{k-1-|v|}v)) \in \delta_2(q, u). \tag{3}$$

That is, we pad rewrites of $M_1$ (from the left) with $k - 1 - |v|$ blank symbols so that the input is reduced by only one symbol for $M_2$. (Note that if $q = q_0$, since $M_1$ has fixed rewrite size, we never pad these lookaheads.) The state $\hat{p}$ indicates that $M_2$ has made a rewrite. There should be no blank symbols for the rest of this cycle (right computation). Therefore if $M_2$ finds a blank symbol while in a hat state, it rejects:

$$\texttt{REJECT} \in \delta_2(\hat{q}, Bx), \quad \text{and} \quad \texttt{REJECT} \in \delta_2(\hat{q}, x_1 B x_2 \$).$$

In subsequent cycles, $M_2$ will delete the blank symbols introduced, one-by-one and *immediately restart*. Unless $M_2$ is in a restarting configuration, it can only delete blank symbols if it is a marked state (i.e., if it guessed that there were blank symbols on the tape at the start of the cycle):

$$\texttt{REWRITE}(x) \in \delta_2(\bar{q}, Bx), \quad \text{deletion of blank symbols in a marked state} \tag{4}$$

$$\texttt{REWRITE}(x) \in \delta_2(q_0, Bx), \quad \text{deletion of blank symbols in the start state.} \tag{5}$$

If $M_2$ reaches the right sentinel in a marked state, and still has no blank symbols under its lookahead, then it rejects (it has verified that its guess about the presence of blank symbols on the tape is incorrect):

$$\texttt{REJECT} \in \delta_2(\bar{p}, u[1, k-1]\$) \quad \forall u[1, k-1] \in \Gamma_1^{k-1}.$$

We have already defined move-right instructions for $M_2$ in state $q_0$. $M_2$ can simulate $M_1$'s move-right steps with only the first symbol under the lookahead. Therefore we can define the rest of $M_2$'s move-right steps simply as follows, for $q' \in \delta_1(q, u)$ and based on just the symbol $u[1]$ of the lookahead (as well as the states $q, q'$). Here, neither $q$ nor $q'$ is the restart state. Also, $x$ does not have the right sentinel as a suffix. If $M_2$ is in a marked state (resp. hat state, state from $Q$) it remains in a marked state (resp. hat state, state from $Q$):

$$\overline{q'} \in \delta_2(\overline{q}, u[1]x), \quad \hat{q}' \in \delta_2(\hat{q}, u[1]x), \quad \text{and} \quad q' \in \delta_2(q, u[1]x).$$

In state $q$ or $\hat{q}$ and with lookahead contents $u$, $M_2$ move rights and rejects (resp. accepts) if in state $q$, $M_1$ moves right and rejects (resp. accepts). Also, it is clear that $\mathcal{L}(M_1) = \mathcal{L}(M_2)$. Moreover, it is easy to see that monotonicity is preserved. □

For the remainder of this paper, we will assume w.l.o.g. that all discussed non-deterministic restarting automata with auxiliary symbols (1) are in RR-semi-det form, (2) carry out move-right steps based on the current state and the first symbol under the lookahead, (3) have fixed rewrite size, and (4) reduce their input by only one symbol per cycle.

## 3 Main Results

For restarting automata with auxiliary symbols and lookahead of size 1, showing that the separation of rewrite and restart step results in an increase in power for these automata. In fact, the result is given for monotone restarting automata also.[2]

**Proposition 5.** *For $X \in \{$(right-left-, left-)mon-,$\lambda\}$,*

$$REG = \mathcal{L}(X\text{-}RWW(1)) \subsetneq \mathcal{L}(right\text{-}left\text{-}mon\text{-}RRWW(1)).$$

*Proof.* Mraz [3] showed that $REG = \mathcal{L}(X\text{-R}(1)) = \mathcal{L}(X\text{-RW}(1)) = \mathcal{L}(X\text{-RWW}(1))$, with $X \in \{$det-mon, det, mon, $\lambda\}$ and this clearly also holds for $X = $ (right-left-, left-)mon. We specify a right-left-mon-RRWW(1) automaton $M$ such that $\mathcal{L}(M) \in LIN - REG$, through the following regular constraints. (Note that $\mathcal{L}(\text{right-left-mon-RRWW}) = LIN$ [2].)

$$(\cent(ab)^*a, b \to \lambda, (cd)^*\$) \quad (\cent(ab)^*a, c \to \lambda, d(cd)^*\$)$$
$$(\cent(ab)^*, a \to \lambda, d(cd)^*\$) \quad (\cent(ab)^*, d \to \lambda, (cd)^*\$) \quad (\cent\lambda\$, \texttt{ACCEPT}).$$

By an enumeration of the left-over context possibilities, it can be shown that $\mathcal{L}(M) = \{(ab)^n(cd)^n \mid n \geq 0\} \cup \{(ab)^{n-1}a(cd)^n \mid n \geq 0\} \cup \{(ab)^{n-1}ad(cd)^{n-1} \mid n \geq 0\} \cup \{(ab)^{n-1}a(cd)^{n-1} \mid n \geq 0\} \in LIN - REG$. □

We can also separate the classes of languages recognised by RWW (RRWW) automata with lookahead 1 from that of those with lookahead 2. The result is also given for monotone restarting automata.

**Proposition 6.** *For all $X \in \{$(right-left, left-)mon, $\lambda\}$,*

$$\mathcal{L}(X\text{-}RWW(2)) - \mathcal{L}(X\text{-}RWW(1)) \neq \emptyset \quad \text{and} \quad \mathcal{L}(X\text{-}RRWW(2)) - \mathcal{L}(X\text{-}RRWW(1)) \neq \emptyset.$$

---

[2] Note that this is only a small improvement on the fact that $\mathcal{L}(X\text{-RWW}(1)) \subsetneq \mathcal{L}(\text{RRWW}(1))$ for all $X \in \{$(right-left-, left-)mon-,$\lambda\}$, which is an immediate consequence of results in [3].

*Proof.* The language $L = \{a^n b^n \mid n \geq 0\}$ is the classic example of a linear language that is not regular. A det-right-left-mon-RWW(2) automaton to recognise $L$ may be specified (deterministically) by the following regular constraints:

$$(¢a^*, ab \rightarrow c, b^*\$), \quad (¢a^*, cb \rightarrow d, b^*\$), \quad (¢a^*, ad \rightarrow c, b^*\$), \quad (¢\lambda\$, \texttt{ACCEPT}), \quad (¢d\$, \texttt{ACCEPT}).$$

On the other hand, no restarting automaton $M$ with just size 1 lookahead can recognise this language, for after the first deletion, the tape contents contain a string not in $\mathcal{L}(M)$, which is excluded by the correctness preserving property. □

It turns out that further separation of language classes for RRWW is not possible. This is the main result of this paper, given in Theorem 7 and Corollary 12.

**Theorem 7.** *For $k \geq 2$ and $X \in \{(\textit{right-left-, left-})\textit{mon}, \lambda\}$, we have*

$$\mathcal{L}(X\textit{-RRWW}(k)) = \mathcal{L}(X\textit{-RRWW}(k+1)).$$

*Proof of Theorem 7.* Assume $M_1 = (Q_1, \Sigma, \Gamma_1, ¢, \$, q_0, k+1, \delta_1)$ is an RRWW(k+1) automaton. We construct $M_2 = (Q_2, \Sigma, \Gamma_2, ¢, \$, q_0, k, \delta_2)$ an RRWW(k) automaton to simulate $M_1$, such that $\mathcal{L}(M_1) = \mathcal{L}(M_2)$.

For this construction, the nondeterminacy of $M_2$ is essential. $M_2$'s lookahead is one symbol shorter than $M_1$'s. So, $M_2$ will simulate $M_1$'s rewrites by guessing the contents of the tape square, $\tau_R$, following the last symbol of its lookahead, contained in tape square $\tau_L$. It will verify this guess within up to one step (of the same cycle), using a compound state holding this information, leaving behind in the compound symbol $\tau_L$, how $M_2$ should read the guessed contents of $\tau_R$ in subsequent cycles; we'll call this instruction $I$. If there is a rewrite starting in $\tau_R$ in a subsequent cycle, $C_i$, then $M_2$ will record in $\tau_R$ that it should ignore $I$ in all cycles after $C_i$. Using the Matching Lemma (Lemma 11) concerning the "interaction" of information in $\tau_L$ and $\tau_R$, $M_2$ will be able to determine which message is most up-to-date. Note that this simulation could not work for $k = 1$, because then $M_1$ can only delete.

We now give the formal proof of the Theorem.

**Notation for $M_2$'s Work Tape.** Let $\Theta_{t,\mathcal{C}} = \pi_{i_{-1}} \pi_{i_0} \pi_{i_1} \pi_{i_2} \cdots \pi_{i_{n-m}} \pi_{i_{n-m+1}} \pi_{i_{n-m+3}}$ denote $M_2$'s work tape at time $t$ in cycle $C_m$ $(m \geq 1)$ of computation $\mathcal{C}$ on an initial input of length $n$, where each $\pi_{i_j}$ is a tape square boundary, for $j \in \{-1, 0\} \cup [n - m + 3]$. Further, with respect to $\Theta_{t,\mathcal{C}}$, we let $\tau_R(\pi_{i_j}, t)$ denote the contents of tape square to the right of $\pi_{i_j}$ at time $t$ (if it exists) and $\tau_L(\pi_{i_j}, t)$ the contents of the tape square to the left of $\pi_{i_j}$ at time $t$ (if it exists). So, we always have, for example, $\tau_R(\pi_{-1}, t) = ¢ = \tau_L(\pi_0, t)$. We call a tape square boundary *internal* if it is between two tape squares. With each cycle, one tape square and boundary are destroyed and for this proof, we say that the second tape square involved in the redex and its boundary to the left are destroyed in the rewrite of the cycle.

**Verification Information and Rewrite Instruction Set Notation.** By *verification information*, VerInf, we will just mean some member of the set of $M_1$'s rewrites, or the special blank symbol, $\mathbf{B} \notin \Gamma_2$, and we will denote the set of verification information as

$$\Pi := \{(q, u[1, k+1], v[1, k], q') \mid (q', \texttt{REWRITE}(v[1, k])) \in \delta_1(q, u[1, k+1])\} \cup \{\mathbf{B}\}.$$

We'll also refer to $\Pi_1 := \Pi - \{\mathbf{B}\}$ as the set of $M_1$'s rewrites. For $\rho = (q, u[1, k+1], v[1, k], q') \in \Pi_1$, we denote to the components of $\rho$ as follows:

$$\texttt{redex}(\rho) := u, \quad \texttt{reduct}(\rho) := v, \quad \texttt{from\_state}(\rho) : q, \quad \text{and} \quad \texttt{to\_state}(\rho) := q'.$$

So, for example, $\texttt{reduct}(\rho)[k+1] = u[k+1]$ and $\texttt{redex}(\rho)[k] = v[k]$. Finally, we denote by $\Pi_2$, the set of $M_2$'s rewrites,

$$\Pi_2 := \{(q, x[1,k], y[1, k-1], q') \mid (q', \texttt{REWRITE}(y[1, k-1])) \in \delta_2(q, x[1,k])\}$$

which will be defined shortly.

**$M_2$'s Tape Alphabet.** $M_2$ has tape alphabet $\Gamma_2 := \Gamma_1 \cup \Delta$, where

$$\Delta := \{(x, \texttt{VerInf}, c_1, c_2) \mid x \in \Gamma_1, \texttt{VerInf} \in \Pi, c_1, c_2 \in \{0, 1, \texttt{neutral}\}\}.$$

The second through fourth components of the information from these compound symbols in $\Delta$ are used for verifying rewrite guesses, updating tape contents, and determining whether updating is necessary.

If $\texttt{VerInf} = \mathbf{B}$, we say that $\texttt{VerInf}$ *is blank*; we refer to the set of compound symbols with blank verification information as $\Delta_B$. Also, we refer to the set of compound symbols with the last component, $c_2$, not equal to $\texttt{neutral}$ as $\Delta_{01}$.

$M_2$ uses compound symbols as either the last and possibly also the first symbol of a reduct. The information $\texttt{VerInf}$ is used for verifying rewrite guesses and updating tape contents; this component will be non-blank in the last symbol of a reduct. $\texttt{VerInf}$ represents the latest simulated rewrite introducing a compound symbol in the tape square as the last symbol of the reduct.

The last two components of the 4-tuples in $\Delta$ take values that help determine when verification information is out of date; the third component gives instructions about information in the following tape square and the fourth component gives instructions about information in the preceding tape square. Their usage will be made precise in Remark 8 and in the description of $M_2$'s rewrite and move-right instructions.

To refer to the different components of compound symbols $z = (z', \texttt{VerInf}, c_1, c_2) \in \Delta$, we introduce the notation $\texttt{comp}_i(z), i \in \{2, 3, 4\}$, which refers to the $i$th component of $z$. On the other hand, $\texttt{comp}_1$ is defined as a homomorphism $\texttt{comp}_1 : \Gamma_2 \cup \{\mathcal{c}, \$\} \to \Gamma_1 \cup \{\mathcal{c}, \$\}$ as follows, for $z \in \Gamma_2 \cup \{\mathcal{c}, \$\}$

$$\texttt{comp}_1(z) := \begin{cases} z & \text{if } z \in \Gamma_1 \cup \{\mathcal{c}, \$\} \\ x & \text{if } z = (x, \texttt{VerInf}, c_1, c_2) \in \Delta. \end{cases}$$

Then we extend $\texttt{comp}_1$ in the natural way to $\texttt{comp}_1 : (\Gamma_2 \cup \{\mathcal{c}, \$\})^* \to (\Gamma_1 \cup \{\mathcal{c}, \$\})^*$.

Further, we inductively define a mapping $h : (\Gamma_2 \cup \{\lambda, \mathcal{c}\}) \times (\Gamma_2 \cup \{\mathcal{c}, \$\})^* \to (\Gamma_1 \cup \{\mathcal{c}, \$\})^*$ by

$$h(z', z) = \begin{cases} \texttt{comp}_1(z) & \text{if } z' \in \Gamma_1 \cup \Delta_B \cup \{\mathcal{c}\}, \text{ or} \\ & \text{if } z' \in \Delta - \Delta_B, z \in \Delta_{01}, \text{ and } \texttt{comp}_4(z) = \texttt{comp}_3(z'), \text{ or} \\ & \text{if } z' = \lambda. \\ \texttt{reduct}(\texttt{comp}_2(z'))[k] & \text{otherwise.} \end{cases}$$

Then we let $h(z', z\alpha) := h(z', z)h(z, \alpha)$, where $z$ is a single symbol.

Since compound symbols may have various components in common, we will sometimes speak of components being introduced into tape squares. If at time $t$ a tape square $\tau$ holds compound symbol $z$ with some component $\texttt{comp}_i(z)$, but at time $t - 1$, $\tau$'s contents held some symbol $z' \in \Gamma_2$ without the same component—that is, either $z' \in \Gamma_1$ or $\texttt{comp}_i(z') \neq \texttt{comp}_i(z)$—then we say that $\texttt{comp}_i(z)$ was *introduced* (into tape square $\tau$) at time $t$.

**$M_2$'s State Set.** For the definition of $Q_2$, we first define the two-by-two mutually exclusive sets $Q_{21}$ and $Q_{22}$ (which are also each mutually exclusive with $Q_1$).

$$Q_{21} := \{(q, \text{VerInf}, c, d, e) \mid q \in Q_1 - \{\text{ACCEPT}, \text{REJECT}\}, \quad \text{VerInf} \in \Pi,$$
$$c, e \in \{0, 1, \text{neutral}\}, d \in \{\text{verify}, \text{ignore}, \text{neutral}\}\}$$

$$Q_{22} := \{q_{u[1,k]} \mid q \in Q_1, u[1,k] \in (\Gamma_1 \cup \{¢\})^k \text{ and } \delta_1(q, u[1,k]\$) \in \{\text{ACCEPT}, \text{REJECT}\}\}$$

$M_2$ has the state set $Q_2 := Q_1 \cup Q_{21} \cup Q_{22}$, where $Q_{22}$ is the set of all possible contexts leading to an accept state for $M_1$, used on exactly the accept step in $M_2$'s computations. The compound states (from $Q_{21}$) are only used to "pick up" information from compound symbols.

To refer the different components of compound symbols $q = (q', \text{VerInf}, c, d, e) \in Q_{21}$, we introduce the notation $\text{COMP}_i(q), i \in \{2, 3, 4, 5\}$, which refers to the $i$th component of $q$. We further define the homomorphism $\text{COMP}_1(q) : Q_2 \to Q_1$ as follows, for $q \in Q_2$.

$$\text{COMP}_1(q) := \begin{cases} q & \text{if } q \in Q_1 \\ p & \text{if } q = p_{u[1,k]} \in Q_{22} \\ p & \text{if } q = (p, \text{VerInf}, c, d, e) \in Q_{21}. \end{cases}$$

Using the mapping $h$ above, we define another mapping $g : Q_2 \times (\Gamma_2 \cup \{¢, \$\})^* \to (\Gamma_1 \cup \{¢, \$\})^*$ by

$$g(q, z) = \begin{cases} \text{comp}_1(z) & \text{if } q \in Q_1, \text{ or} \\ & \text{if } z \in \Delta_{01}, \text{ and } \text{comp}_4(z) = \text{COMP}_3(q), \text{ or} \\ & \text{if } z = \{¢, \$\}. \\ \text{reduct}(\text{COMP}_2(q))[k] & \text{otherwise.} \end{cases}$$

Then we let $g(q, z\alpha) := g(q, z)h(z, \alpha)$, where $z$ is a single symbol.

The presentation of the proof is somewhat eased by first presenting some guiding properties for $M_2$ that the definition of rewrite and move-right steps will have to obey; this is the purpose of Remark 8 (some comments on Remark 8 follow). After this, we will prove some facts about $M_2$ based on these properties and use these results in the remainder of our definition of $M_2$ that follows.

**Remark 8.** *$M_2$ will be defined according to the six following invariants:*

*(I1) $M_2$'s rewrites will be of the form $(p, \text{REWRITE}(y[1, k-1])) \in \delta_2(q, x[1, k])$ where:*

  *(a) The last symbol of the reduct, $y[k-1]$, is from $\Delta - \Delta_B$ and is such that $\text{comp}_2(y[k-1]) \in \Pi_1$ is the rewrite of $M_1$ simulated.*

  *(b) The first symbol of the reduct, $y[1]$, is from $\Delta_{01} \cup \Gamma_1$.*

  *(c) All remaining symbols of the reduct, $y[i], i \in \{2, \ldots, k-2\}$ are from $\Gamma_1$.*

*(I2) $M_2$ will only write a symbol from $\Delta_{01}$ if in a compound state. In particular, if $M_2$ is in compound state $q$ and writes symbol $y \in \Delta_{01}$, then $\text{comp}_2(y) = \text{COMP}_2(q)$ and $\text{comp}_4(y) = \text{COMP}_3(q)$.*

*(I3) $M_2$ will always enter a compound state after carrying out a rewrite step. In fact, if $M_2$ is in compound state $q$ after writing compound symbol $y[k-1] \in \Delta - \Delta_B$, then $\text{COMP}_2(q) = \text{comp}_2(y[k-1])$, $\text{COMP}_3(q) = \text{comp}_3(y[k-1])$, $\text{COMP}_4(q) \in \{\text{verify}, \text{ignore}\}$, and if $x[k] \in \Delta - \Delta_B$, then $\text{COMP}_5(q) = \text{comp}_3(x[k])$, otherwise $\text{COMP}_5(q) = \text{neutral}$.*

*(I4) $M_2$ enters a compound state after reading a compound symbol from $\Delta - \Delta_B$ as the first symbol under the lookahead. Otherwise, after a move-right step $M_2$ must be in a state from $Q_1$. In fact, if $M_2$ reads symbol $z \in \Delta$, then it enters a compound state $q$ such that $\text{COMP}_2(q) = \text{comp}_2(z)$, $\text{COMP}_3(q) = \text{comp}_3(z)$, and $\text{COMP}_4(q) = \text{COMP}_5(q) = \text{neutral}$.*

(I5) $M_2$ in compound state $q$ with $\texttt{COMP}_4(q) \in \{\texttt{verify}, \texttt{ignore}\}$ rejects if it reads a compound symbol $z \in \Delta$ such that $\texttt{COMP}_3(q) = \texttt{comp}_4(z)$.

Moreover, if $M_2$ does not reject, then

(a) if $\texttt{COMP}_4(q) = \texttt{verify}$, then $M_2$ checks that $\texttt{reduct}(\texttt{COMP}_2(q))[k+1] = \texttt{comp}_1(z)$ ($M_2$ verifies the symbol currently scanned). Furthermore, if $\texttt{COMP}_5(q) \in \{0, 1\}$ and $z \in \Delta_{01}$, $M_2$ also assures that $\texttt{COMP}_5(q) = \texttt{comp}_4(z)$ ($M_2$ verifies that the currently scanned symbol holds the most up-to-date information).

(b) if $\texttt{COMP}_4(q) = \texttt{ignore}$, $\texttt{COMP}_5(q) \in \{0, 1\}$, and $z \in \Delta_{01}$, then $M_2$ assures that $\texttt{COMP}_5(q) \neq \texttt{comp}_4(z)$ ($M_2$ verifies that the information in the currently scanned symbol, $z$, is out-of-date).

Then $M_2$ (in both cases of $\texttt{COMP}_4(q)$) enters some state $p$ such that $\texttt{COMP}_1(q) = \texttt{COMP}_1(p)$ and if $p \notin Q_1$, then $\texttt{COMP}_4(p) = \texttt{COMP}_5(p) = \texttt{neutral}$ and $\texttt{COMP}_i(p) = \texttt{comp}_i(z)$ for $i \in \{2, 3\}$.

(I6) Let $p \in Q_2 - (\{\texttt{ACCEPT, REJECT}\} \cup Q_{22})$.

(a) There is some left computation on prefix $\textcent\alpha \in \Gamma_2^*$ in which $M_2$ reaches state $p$ if and only if there is some left computation on prefix $h(\lambda, \textcent\alpha)$ that puts $M_1$ in state $q = \texttt{COMP}_1(p)$.

(b) There is some right computation on prefix[3] $z\alpha$ after which $M_2$ enters state $p$ where $z \in \Gamma_2, \alpha \in \Gamma_2^*$ starting in state $p'$ if and only if there is some right computation on prefix $h(z, \alpha)$ after which $M_1$ enters state $\texttt{COMP}_1(p)$ starting in state $\texttt{COMP}_1(p')$.

(I1-I3) concern rewrite steps, (I4-I5) concern move-right steps, and (I6) is the main statement that ensures this proof works (valid simulations).

(I4) ensures that $M_2$ can update tape contents after reading a compound symbol from $\Delta$, but that it should not verify that the rewrite guess indicated in this information is correct ($\texttt{COMP}_4(q) = \texttt{neutral}$). In fact, this verification should have taken place directly following the rewrite (in the same cycle) as is indicated in (I3) ($\texttt{COMP}_4(q) \in \{\texttt{verify, ignore}\}$). Points (I3-I5) together indicate that $M_2$ can only be in a state with fourth component equal to a member of $\{\texttt{verify, ignore}\}$ at most once in a cycle: verification of the rewrite guess happens during a single move-right step in the same cycle. By the same token, $M_2$ can only be in a state with fifth component non-equal to $\texttt{neutral}$ during the same single move-right step of the cycle: verification of the updated-ness of the last symbol under the lookahead can happen only in the step after a rewrite, since move-right steps are only defined with respect to the first symbol under the lookahead.

(I2) ensures that $M_2$ can detect when an update of the tape contents has been written onto the tape. (I5) permits $M_2$ to keep track of cycle orders, to the extent that is necessary here. (See Lemma 11.)

From Remark 8, we easily obtain the following three facts:

**Lemma 9.** *At no time $t$ in $M_2$'s computation $\mathcal{C}$ is there an interior square boundary $\pi$ on $M_2$'s work tape $\Theta_{t,\mathcal{C}}$ such that $\tau_L(\pi, t) \in \Gamma_1 \cup \Delta_B \cup \{\textcent\}$ and $\tau_R(\pi, t) \in \Delta_{01}$. (No symbol from $\Gamma_1 \cup \Delta_B \cup \{\textcent\}$ directly precedes a symbol from $\Delta_{01}$ on $M_2$'s work tape at any time $t$ in the computation.)*

*Proof.* This follows from (I1-I4). □

**Corollary 10.** *$M_2$ cannot read a symbol from $\Delta_{01}$ in a state from $Q_1$.*

The following Matching Lemma shows that $M_2$ can detect the order of rewrites over consecutive tape squares.

---
[3] By *prefix in a right computation* we mean the prefix of the segment of work tape contents following the rewrite.

**Lemma 11** (Matching Lemma). *At time $t$ in $M_2$'s computation $\mathcal{C}$ let $\pi$ be an interior tape square boundary on $M_2$'s work tape $\Theta_{t,\mathcal{C}}$. Suppose $\tau_L(\pi, t) \in \Delta - \Delta_B$ and $\tau_R(\pi, t) \in \Delta_{01}$. Then there are two cycles $C_{j_1}, C_{j_2} \in \mathcal{C}$, such that*

1. $M_2$ uses rewrite $\rho_i = (q_i, x_i[1,k], y_i[1,k-1], q_i')$ at time $t_i$ in $C_{j_i}$ ($i \in [2]$) such that $C_{j_1}$ introduced $\texttt{comp}_1(\tau_L(\pi, t)) = \texttt{comp}_1(y_1[k-1])$, and $C_{j_2}$ introduced $\texttt{comp}_4(\tau_R(\pi, t)) = \texttt{comp}_4(y_2[1]) \in \{0, 1\}$.

2. (a) $\texttt{comp}_3(\tau_L(\pi, t_1)) = \texttt{comp}_4(\tau_R(\pi, t_2))$, implies $t_1 < t_2$.
   (b) $\texttt{comp}_3(\tau_L(\pi, t_1)) \neq \texttt{comp}_4(\tau_R(\pi, t_2))$, implies $t_1 > t_2$.

*Proof.* (1) follows from (I1). (2a) follows from (I2) and (I4). (2b) follows from (I3) and (I5). □

In case (2a) of the Matching Lemma, $M_2$ should update the tape square (in memory) $\tau_R(\pi, t)$ as it reads it, and in case (2b), $M_2$ should ignore the instruction in $\tau_L(\pi, t)$ to update the information in $\tau_R(\pi, t)$, since it is now "out of date". We also remark that the Matching Lemma helped provide the definition of the mappings $h$ and $g$.

We now describe the rewrite and move-right instruction for $M_2$ with $k > 2$. The case for $k = 2$ is easily obtained from this by merging the requirements for the first and last symbols in reducts of the case $k > 2$.

**Rewrite steps of $M_2$.** Let $\rho = (q, u[1, k+1], v[1, k], q') \in \Pi_1$. We define a set of $M_2$'s rewrites required for simulating $\rho$ of the form

$$\rho' = (p, x[1, k], y[1, k-1], p') \subseteq \Pi_2$$

with the following component requirements.

1. $p = q$ if $p \in Q_1$, and $p = (q, \rho'', \texttt{comp}_3(\tau_L(\pi, t)), \texttt{neutral}, \texttt{neutral})$, otherwise, where $\rho''$ has further constraints with respect to $x[1]$. (See Item (7).)

2. For $p'$, we have (by (I3))

$$p' = \begin{cases} (q', \rho, \texttt{comp}_3(y[k-1]), \texttt{verify}, \texttt{neutral}) & \text{if } x[k] \in \Gamma_1 \cup \Delta_B, \\ (q', \rho, \texttt{comp}_3(y[k-1]), \texttt{ignore}, \texttt{comp}_3(x[k])) & \text{if } x[k] \in \Delta - \Delta_B \text{ and only in (6b)}, \\ (q', \rho, \texttt{comp}_3(y[k-1]), \texttt{verify}, \texttt{comp}_3(x[k])) & \text{if } x[k] \in \Delta - \Delta_B \text{ and only in (6a)}. \end{cases}$$

3. Any $x[2, k-1] \in \Gamma_2^{k-2}$ such that $h(x[1], x[2, k-1]) = u[2, k-1]$.

4. $y[2, k-2] = v[2, k-2]$.

5. $y[k-1] = (v[k-1], \rho, c_1, \texttt{neutral})$, with $c_1 \in \{0, 1\}$, by (I1).

6. (a) any $x[k] \in \Gamma_2$ such that $h(x[k-1], x[k]) = u[k]$, or
   (b) any $x[k] \in \Delta - \Delta_B$ such that $\texttt{reduct}(\texttt{comp}_2(x[k]))[k] = u[k+1]$, and $\texttt{comp}_3(x[k]) \in \{0, 1\}$.

7. Finally for $x[1], y[1]$,

   - If $p \in Q_1$, then $y[1] = v[1]$ and any $x[1] \in \Gamma_2 \cup \{\texttt{¢}\}$ such that $\texttt{comp}_1(x[1]) = u[1]$ will suffice.
   - If $p \in Q_{21}$, then $y[1] = (v[1], \mathbf{B}, \texttt{neutral}, \texttt{COMP}_3(p))$ and
     - any $x[1] \in (\Gamma_2 \cup \{\texttt{¢}\}) - \Delta_{01}$ such that $\texttt{comp}_1(x[1]) = \texttt{redex}(\texttt{COMP}_2(p))[k+1]$ and $\texttt{reduct}(\texttt{COMP}_2(p))[k] = u[1]$, or

- any $x[1] \in \Delta_{01}$ such that
  * $\texttt{COMP}_3(p) \neq \texttt{comp}_4(x[1])$, $\texttt{comp}_1(x[1]) = \texttt{redex}(\texttt{COMP}_2(p))[k+1]$ and $\texttt{reduct}(\texttt{COMP}_2(p))[k] = u[1]$, or
  * $\texttt{COMP}_3(p) = \texttt{comp}_4(x[1])$ and $\texttt{comp}_1(x[1]) = u[1]$.

  by the Matching Lemma.

There are no other rewrites in $\delta_2$.

Note that $M_2$ cannot rewrite over the right sentinel, since it always simulates $M_1$'s rewrites using only the first $k$ symbols and $M_1$ has fixed rewrite size.

**Move-right steps of $M_2$ *not* derived from $M_2$'s move-right steps.** We suppose without loss of generality that $M_1$ doesn't rewrite over the right sentinel and then immediately halt. There are two types of move-right steps for $M_2$ that are not derived from $M_1$'s move-right steps, for verifying rewrite guesses; they are therefore derived from $M_1$'s rewrites. These two cases, for $\delta_2(p, x[1,k])$ are when $p \in Q_{21}$ with $\texttt{COMP}_4(p) \in \{\texttt{verify}, \texttt{ignore}\}$. In these move-right steps, $M_2$ simply verifies that Invariant (I5) is maintained and, if so, moves right and into state

$$\begin{cases} \texttt{COMP}_1(p) & \text{if } x[1] \in \Gamma_1, \text{ and} \\ (\texttt{COMP}_1(p), \texttt{comp}_2(x[1]), \texttt{comp}_3(x[1]), \texttt{neutral}, \texttt{neutral}) & \text{otherwise,} \end{cases}$$

indicating that $M_2$ remains in the "same" state (with respect to $M_1$'s state), picks up $x[1]$'s verification information (in case it must update tape contents), and its matching information (to keep track of the order of rewrites). The fourth and fifth components are always $\texttt{neutral}$ in the compound state following any step that does not verify a rewrite step.

**Move-right steps of $M_2$ derived from $M_2$'s move-right steps.** Other than the above described move-right steps, $M_2$'s move-right steps nondeterministically simulate those of $M_1$ simultaneously updating tape contents because of rewrite guesses. Recall that since $M_1$ is in the RR-semidet-form, we only need to consider the first symbol under the lookahead for $M_1$'s move-right steps (so, in particular, we can talk about move-right steps in $\delta_1$ on a lookahead contents of size $k$ instead of $k+1$).

Let
$$q' \in \delta_1(q, u[1, k+1]) \tag{6}$$
be a move-right step for $M_1$.

Firstly, $q' \in \delta_2(q, u[1]x[2,k])$, for all $h(x[1], x[2,k]) = u[2,k]$.

In addition, $M_2$ has the following instructions:

If $M_1$ accepts/rejects/restarts with less than $k + 1$ symbols under the lookahead, then so can $M_2$; that is, if $\delta_1(q, u[1, j]) = \texttt{ACCEPT}$ (resp. REJECT, RESTART) for $1 \leq j < k$, with $u[j] = \$$, then $\delta_2(p, x[1, j]) = \texttt{ACCEPT}$ (resp. REJECT, RESTART) with $x[j] = \$$ and such that $\texttt{COMP}_1(p) = q$, and for all $x[1, j-1] \in (\Gamma_2 \cup \{\texttt{¢}\}) \cdot \Gamma_2^{j-2}$ such that $g(q, x[1, j-1]) = u[1, j-1]$.

In the remaining description, we describe the simulation move-right steps in which $M_1$ always has $k + 1$ symbols under the lookahead.

If $q' = \texttt{ACCEPT}$ (so $u[k+1] = \$$), then we have, for $q_{u[1,k]} \in Q_{22}$, $q_{u[1,k]} \in \delta_2(p, x[1, k])$, and

$$\delta_2(q_{u[1,k]}, x[2,k]z) \ni \begin{cases} \texttt{ACCEPT} & \text{if } z = \$, \text{ and} \\ \texttt{REJECT} & \text{otherwise.} \end{cases}$$

for all $p$ such that $\texttt{COMP}_1(p) = q$ and $\texttt{COMP}_4(p) = \texttt{COMP}_5(p) = \texttt{neutral}$, and for all $z \in \Gamma_2$, and for all $x[1, k] \in (\Gamma_2 \cup \{\texttt{¢}\}) \cdot \Gamma_2^{k-1}$ such that $g(p, x[1, k]) = u[1, k]$. Here, $M_2$ first guesses that $M_1$ would



accept and then verifies its guess. We must have $\texttt{COMP}_4(p) = \texttt{COMP}_5(p) = \texttt{neutral}$, because in the step after rewriting we have assumed that $M_1$ does not immediately halt after rewriting.

If $q' = \texttt{REJECT}$, then we have simply $\texttt{REJECT} \in \delta_2(p, x[1, k])$ for all $p$ such that $\texttt{COMP}_1(p) = q$ and for all $x[1, k] \in (\Gamma_2 \cup \{\cent\}) \cdot \Gamma_2^{k-1}$ such that $g(p, x[1, k]) = u[1, k]$, so long as $\texttt{COMP}_4(p) = \texttt{COMP}_5(p) = \texttt{neutral}$. $M_2$ can guess that the $M_1$ would reject; if this is not the case, there is still some computation that does not reject.

By Corollary 10, the remaining cases for the simulation of (6) are where $M_2$ reads a compound symbol (as the first symbol under the lookahead) and/or is in a compound state.

Suppose $p \in Q_1$, i.e., $p = q$. By Corollary 10, we must have $x[1] \in \Delta - \Delta_{01}$ and therefore $\texttt{comp}_1(x[1]) = u[1]$. Now $M_2$ simply picks up the information in $x[1]$ and moves right as $M_1$ would:

$$(q', \texttt{comp}_2(x[1]), \texttt{comp}_3(x[1]), \texttt{neutral}, \texttt{neutral}) \in \delta_2(p, x[1, k]). \tag{7}$$

Finally, suppose $p \in Q_{21}$; then $\texttt{COMP}_1(p) = q$. The only case left to treat is where $\texttt{COMP}_4(p) = \texttt{neutral}$.

1. If $x[1] \in (\Gamma_2 \cup \{\cent\}) - \Delta_{01}$, then $\texttt{comp}_1(x[1]) = \texttt{redex}(\texttt{COMP}_2(p))[k+1]$ and $\texttt{reduct}(\texttt{COMP}_2(p))[k] = u[1]$.

2. If $x[1] \in \Delta_{01}$. Then by the Matching Lemma,
   (a) $\texttt{COMP}_3(p) \neq \texttt{comp}_4(x[1])$, $\texttt{comp}_1(x[1]) = \texttt{redex}(\texttt{COMP}_2(p))[k+1]$ and $\texttt{reduct}(\texttt{COMP}_2(p))[k] = u[1]$, or
   (b) $\texttt{COMP}_3(p) = \texttt{comp}_4(x[1])$ and $\texttt{comp}_1(x[1]) = u[1]$.

$M_2$ rejects for all other contexts (except where it can rewrite).

$M_2$'s rewrite and move-right steps being entirely determined by $M_1$'s, it follows that $\mathcal{L}(M_1) = \mathcal{L}(M_2)$. $\square$

As a corollary of Theorem 7, we have the following lookahead hierarchy collapsal.

**Corollary 12.** *For $k \geq 2$ and $X \in \{$(left-, right-left-)mon, $\lambda\}$, we have*

$$\mathcal{L}(X\text{-}RRWW) = \bigcup_{k=2}^{\infty} \mathcal{L}(X\text{-}RRWW(k)) = \mathcal{L}(X\text{-}RRWW(2))$$

Corollary 12 reduces the most important question concerning restarting automata—whether the separation of rewrite and restart steps results in an increase in power—to the same question about restarting automata with lookahead length 2: $\mathcal{L}(RWW) = \mathcal{L}(RRWW) \iff \mathcal{L}(RWW) = \mathcal{L}(RRWW(2))$. Theorem 7 also leads to an improvement on a result of [6] with the following corollary, which was proven for $k \geq 3$ (Corollary 13), as well as a corresponding corollary for right-left-monotonicity (Corollary 14).

**Corollary 13.** *For all $k \geq 2$ and $X \in \{$left-mon, mon$\}$, we have $\mathcal{L}(X\text{-}RRWW(k)) = CFL$.*

**Corollary 14.** *For all $k \geq 2$, we have $\mathcal{L}(right\text{-}left\text{-}RRWW(k)) = LIN$.*

## 4 Concluding Remarks

We showed that the restriction on lookahead length is not as important a restriction for restarting automata with auxiliary symbols as opposed to those without auxiliary symbols, so long as restart and rewrite steps are separated, distinguishing only two different language classes for RRWW automata. The respective question for RWW automata remains open.

**Acknowledgements.** We thank the anonymous reviewers for their helpful comments.

# References

[1] P. Jančar, F. Mráz, M. Plátek, and J. Vogel. On monotonic automata with a restart operation. *Journal of Automata, Languages and Combinatorics*, 4(4):287–312, 1999.

[2] T. Jurdziński, F. Mráz, F. Otto, and M. Plátek. Degrees of non-monotonicity for restarting automata. *Theoretical Computer Science*, 369:1–34, 2006.

[3] F. Mráz. Lookahead hierarchies of restarting automata. *Journal of Automata, Languages and Combinatorics*, 6(4):493–506, 2001.

[4] G. Niemann and F. Otto. Restarting automata and prefix rewriting systems. Technical report, Kassel University, 1999.

[5] F. Otto. Restarting automata. In Z. Esik, C. Martin-Vide, and V. Mitrana, editors, *Recent Advances in Formal Languages and Applications*, volume 25 of *Studies in Computational Intelligence*, pages 269–303. Springer-Verlag, Berlin, 2006.

[6] N. Schluter. On lookahead hierarchies for monotone and deterministic restarting automata with auxiliary symbols (extended abstract). In *Developments in Languages Theory, 14th International Conference, DLT 2010*, London, Ontario, Canada, 2010.